\documentclass{iaus}
\usepackage{graphicx}

\newcommand{\apj}{ Astrophys. J.}
\newcommand{\apjl}{Astrophys. J. Let.}
\newcommand{\apjs}{Astrophys. J. Suppl.}
\newcommand{\aj}{Astron. J.}
\newcommand{\aap}{Astron. \& Astrophys.}
\newcommand{\araa}{Ann. Rev.  Astron. Astrophys.}
\newcommand{\jgr}{J. Geophys. Res.}
\newcommand{\mnras}{Mon. Not. R. Astron. Soc.}
\newcommand{\nat}{Nature}
\newcommand{\pasj}{Pub. Astron. Soc. Jap.}

\newcommand{\icarus}{Icarus}

\title[The debris disk - terrestrial planet connection]
{The debris disk - terrestrial planet connection} 

\author[Raymond et al.]
{Sean~N.~Raymond$^{1,2\ast}$, 
Philip~J.~Armitage$^{3,4}$,
Amaya~Moro-Mart{\' i}n$^{5,6}$,
Mark~Booth$^{7}$,
Mark~C.~Wyatt$^{7}$
John C.~Armstrong$^{8}$,
Avi~M.~Mandell$^{9}$,
\& Franck~Selsis$^{1,2}$}

\affiliation{
$^{1}$Universit{\'e} de Bordeaux, Observatoire Aquitain des Sciences de l'Univers, 2 rue de l'Observatoire, BP 89, F-33271 Floirac Cedex, France\\[\affilskip]
$^{2}$CNRS, UMR 5804, Laboratoire d'Astrophysique de Bordeaux, 2 rue de l'Observatoire, BP 89, F-33271 Floirac Cedex, France\\[\affilskip]
$^{3}$JILA, University of Colorado, Boulder CO 80309, USA\\[\affilskip]
$^{4}$Department of Astrophysical and Planetary Sciences, University of Colorado, Boulder CO 80309, USA\\[\affilskip]
$^{5}$Departamento de Astrofisica, CAB (CSIC-INTA),  Instituto Nacional de Tecnica Aeroespacial, Torrejon de Ardoz, 28850, Madrid,  Spain\\[\affilskip]
$^{6}$Department of Astrophysical Sciences, Princeton University, Peyton Hall, Ivy Lane, Princeton, NJ 08544, USA\\[\affilskip]
$^{7}$Institute of Astronomy, Cambridge University, Madingley Road, Cambridge, UK\\[\affilskip]
$^{8}$Department of Physics, Weber State University, Ogden, UT, USA\\[\affilskip]
$^{9}$NASA Goddard Space Flight Center, Code 693, Greenbelt, MD 20771, USA\\[\affilskip]
$^\ast$E-mail:  rayray.sean@gmail.com.
}

\pubyear{2010}
\volume{276}  
\pagerange{??}
\jname{The Astrophysics of Planetary Systems}
\editors{A. Sozzetti, M. G. Lattanzi, A. P. Boss, eds.}
\begin{document}

\maketitle


\begin{abstract}
The eccentric orbits of the known extrasolar giant planets provide evidence that most planet-forming environments undergo violent dynamical instabilities.  Here, we numerically simulate the impact of giant planet instabilities on planetary systems as a whole.  We find that populations of inner rocky and outer icy bodies are both shaped by the giant planet dynamics and are naturally correlated.  Strong instabilities -- those with very eccentric surviving giant planets -- completely clear out their inner and outer regions.  In contrast, systems with stable or low-mass giant planets form terrestrial planets in their inner regions and outer icy bodies produce dust that is observable as debris disks at mid-infrared wavelengths.   Fifteen to twenty percent of old stars are observed to have bright debris disks (at $\lambda \sim 70 {\mu}m$) and we predict that these signpost dynamically calm environments that should contain terrestrial planets. 
\end{abstract}

\firstsection 
\section{Introduction}

Circumstellar disks of gas and dust orbiting young stars are expected to produce three broad classes of planets in radially-segregated zones (e.g., Kokubo \& Ida 2002).  The inner disk forms terrestrial (rocky) planets because it contains too little solid mass to rapidly accrete giant planet cores.  Terrestrial planets form in 10-100 million years (Myr) via collisional agglomeration of Moon- to Mars-sized planetary embryos and a swarm of 1-10$^3$~km sized planetesimals (Chambers 2004; Kenyon \& Bromley 2006; Raymond et al 2009).  

From roughly a few to a few tens of Astronomical Units (AU), giant planet cores grow and accrete gaseous envelopes if the conditions are right.  Despite their large masses, gas giants must form within the few million year lifetime of gaseous disks (Haisch et al 2001) and be present during the late phases of terrestrial planet growth.  The known extrasolar giant planets have a broad eccentricity distribution (Butler et al 2006) that is quantitatively reproduced if dynamical instabilities occurred in 70-100\% of all observed systems (Chatterjee et al 2008; Juric \& Tremaine 2008; Raymond et al 2010).  The onset of instability may be caused by the changing planet-planet stability criterion as the gas disk dissipates (Iwasaki et al 2001), resonant migration (Adams \& Laughlin 2003), or chaotic dynamics (Chambers et al 2006), leading to a phase of planet-planet scattering and the removal of one or more planets from the system by collision or hyperbolic ejection (Rasio \& Ford 1996; Weidenschilling \& Marzari 1996).  

Finally, in the outer regions of planetary systems the growth time scale exceeds the lifetime of the gas disk, and the end point of accretion is a belt of Pluto-sized (and smaller) bodies (Kenyon \& Luu 1998).  Debris disks trace these icy leftovers of planet formation.  Debris disks consist of warm or cold dust observed around older stars, typically at infrared wavelengths ($\lambda \sim 10-100 \mu m$) and provide evidence for the existence of planetesimals because the lifetime of observed dust particles under the effects of collisions and radiation pressure is far shorter than the typical stellar age, implying a replenishment via collisional grinding of larger bodies (e.g., Wyatt 2008; Krivov 2010).  

Here we perform a large ensemble of N-body simulations to model the interactions between the different radial components of planetary systems: formation and survival of terrestrial planets, dynamical evolution and scattering of giant planets, and dust production from collisional grinding.  By matching the orbital distribution of the giant planets, we infer the characteristics of as-yet undetected terrestrial planets in those same systems.  We calculate the spectral energy distribution of dust in the system by assuming that planetesimal particles represent a population of smaller objects in collisional equilibrium (as in Booth et al 2009).  We then correlate outcomes in the different radial zones and link to two key observational quantities: the orbital properties of giant planets and debris disks.  The reader is referred to Raymond et al (2011a,b) for more details. 

\section{Methods}
Our initial conditions include $9 M_\oplus$ in planetary embryos and planetesimals from 0.5 to 4 AU, three giant planets on marginally stable orbits from Jupiter's orbital distance of 5.2~AU out to $\sim$10~AU (depending on the masses), and an outer 10~AU-wide disk of planetesimals containing 50~$M_\oplus$.  Giant planet masses were drawn randomly according to the observed exoplanet mass function (Butler et al 2006) ${\rm d}N / {\rm d}M \propto M^{-1.1}$ for $M_{\rm Sat} < M < 3 M_{\rm Jup}$.  Each of our 160 fiducial simulations was integrated for 100-200 million years using the {\tt Mercury} (Chambers 1999) hybrid integrator, paying careful attention to the energy conservation of the integrator. We post-process the simulations to compute the spectral energy distribution of dust by treating planetesimal particles as aggregates in collisional equilibrium (Dohnanyi 1969) to calculate the incident and re-emitted flux (Booth et al 2009).  Though certainly oversimplified, these initial conditions generically reproduce the predicted state of a planetary system at the time of the dissipation of the gaseous protoplanetary disk.   

We do not include the effects of planetary migration for several reasons.  First, current migration theories fail to reproduce most characteristics of the known exoplanets (Howard et al 2010).  Second, the effects of giant planet migration on terrestrial planet formation are thought to be weak(Raymond et al 2006; Mandell et al 2007; Fogg \& Nelson 2007) compared with the effect of instabilities (Veras \& Armitage 2006).  Finally, dynamical instabilities are thought to occur when the disk is mostly or completely dissipated, {\em after} any migration has occurred such that the imprint of instabilities on the surviving planets should be far more pronounced.

\section{Results}
An unstable system evolves as follows (Figure~1).  At early times accretion proceeds in the inner disk, the giant planets gravitationally interact, the outer disk is mostly passive, and the only significant interaction between the components is that the giant planets dynamically clear out nearby small bodies. The onset of instability causes a punctual increase in the giant planets' eccentricities and, depending on the details of the scattering event, the orbit of one or more giant planets intrudes into the inner and/or outer disk.  Particles in the vicinity of a scattered giant planet are rapid destabilized to either be ejected from the system or collide with the central star, and this process continues until the instability concludes (usually with the ejection of one or two giant planets; e.g., Raymond et al 2010).  At this point, the surviving small bodies have been shaped by the dynamical instability but their continued evolution is governed by the new dynamical state of the surviving giant planets.  The number and spacing of terrestrial planets that form depends on the eccentricities of the surviving embryos (if any), perturbed both during and after the instability.  These perturbations span a continuous range but the outcome is quantized into a discrete number of terrestrial planets during the accretion process (Levison \& Agnor 2003).  If perturbations are weak -- if the giant planets collide rather than scattering (or are dynamically stable or low-mass) -- then embryos' eccentricities remain small and feeding zones narrow and several terrestrial planets form.  For stronger giant planet perturbations, feeding zones widen and fewer terrestrial planets form, although the total mass in planets tends to decrease because stronger perturbations imply that the giant planets were scattered closer to the terrestrial planet region so more embryos end up on unstable orbits.  In systems where embryos' radial excursions are comparable to the radial extent of the surviving disk only one planet forms, usually on an excited orbit.  In the simulation from Figure~1, the lone terrestrial planet did not accrete from a disk of excited embryos but rather was the only planet to {\em survive} the instability.  Perturbations during, not after, the instability determined the outcome.  The outer disk's evolution is also governed by giant planet perturbations: icy planetesimals that survive the instability are subject to secular forcing that determines the collisional timescale and the rate of dust production.

\begin{figure}
\includegraphics[width=0.63\textwidth]{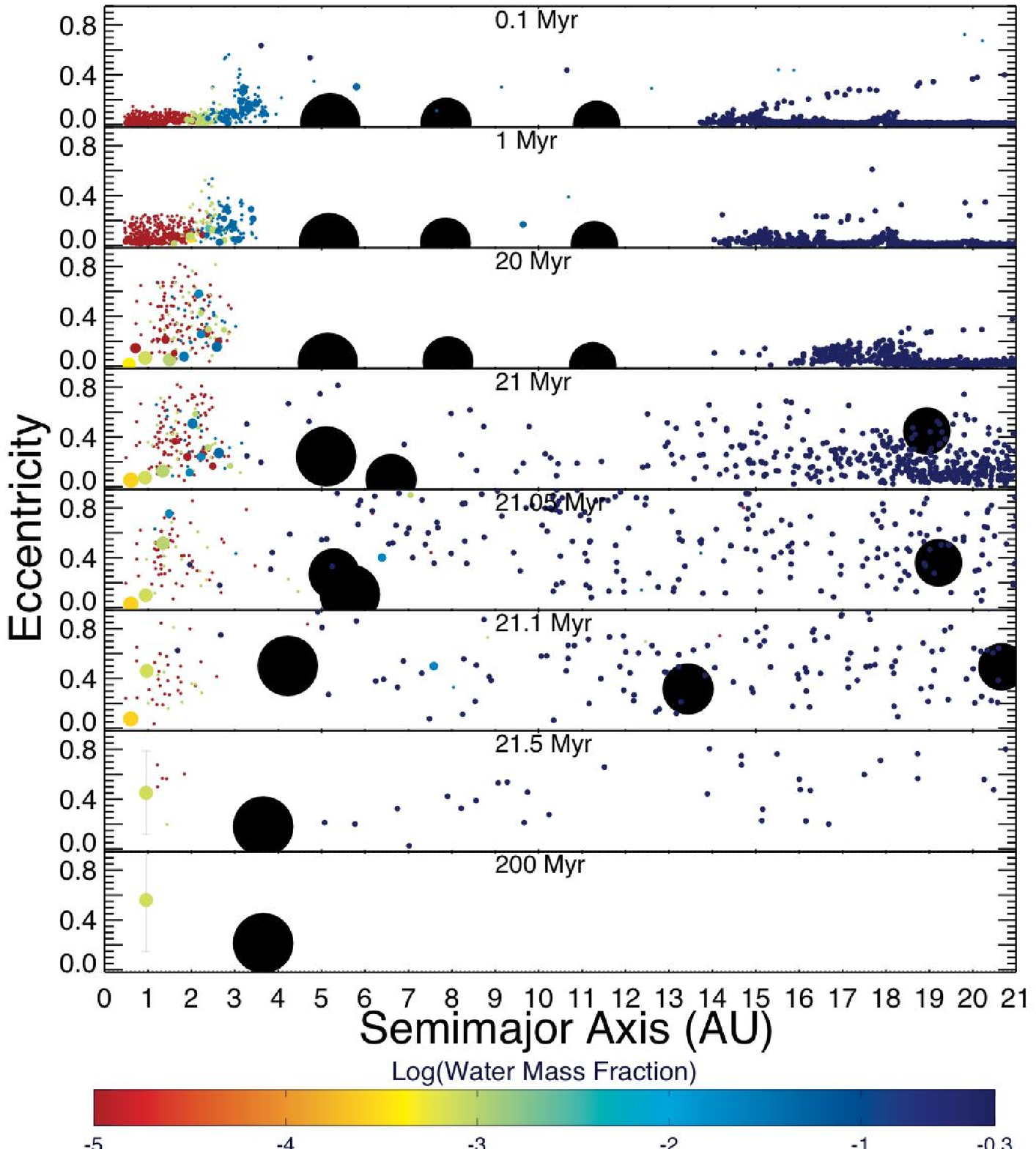}
\vskip -3.7in
\hspace {0.63\textwidth}
\includegraphics[width=0.37\textwidth]{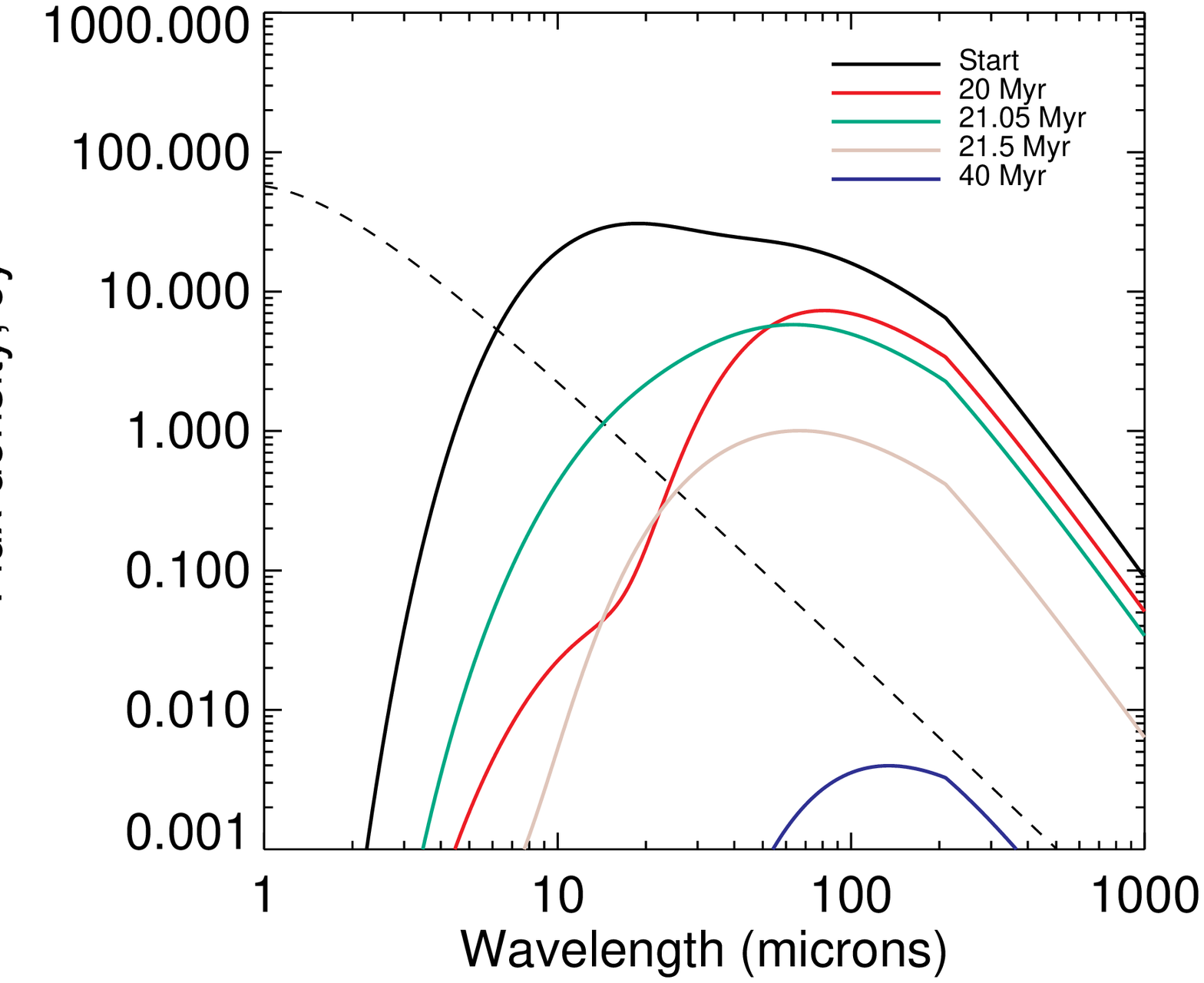}
\vskip 0.01in 
\hspace {0.63\textwidth} 
\includegraphics[width=0.37\textwidth]{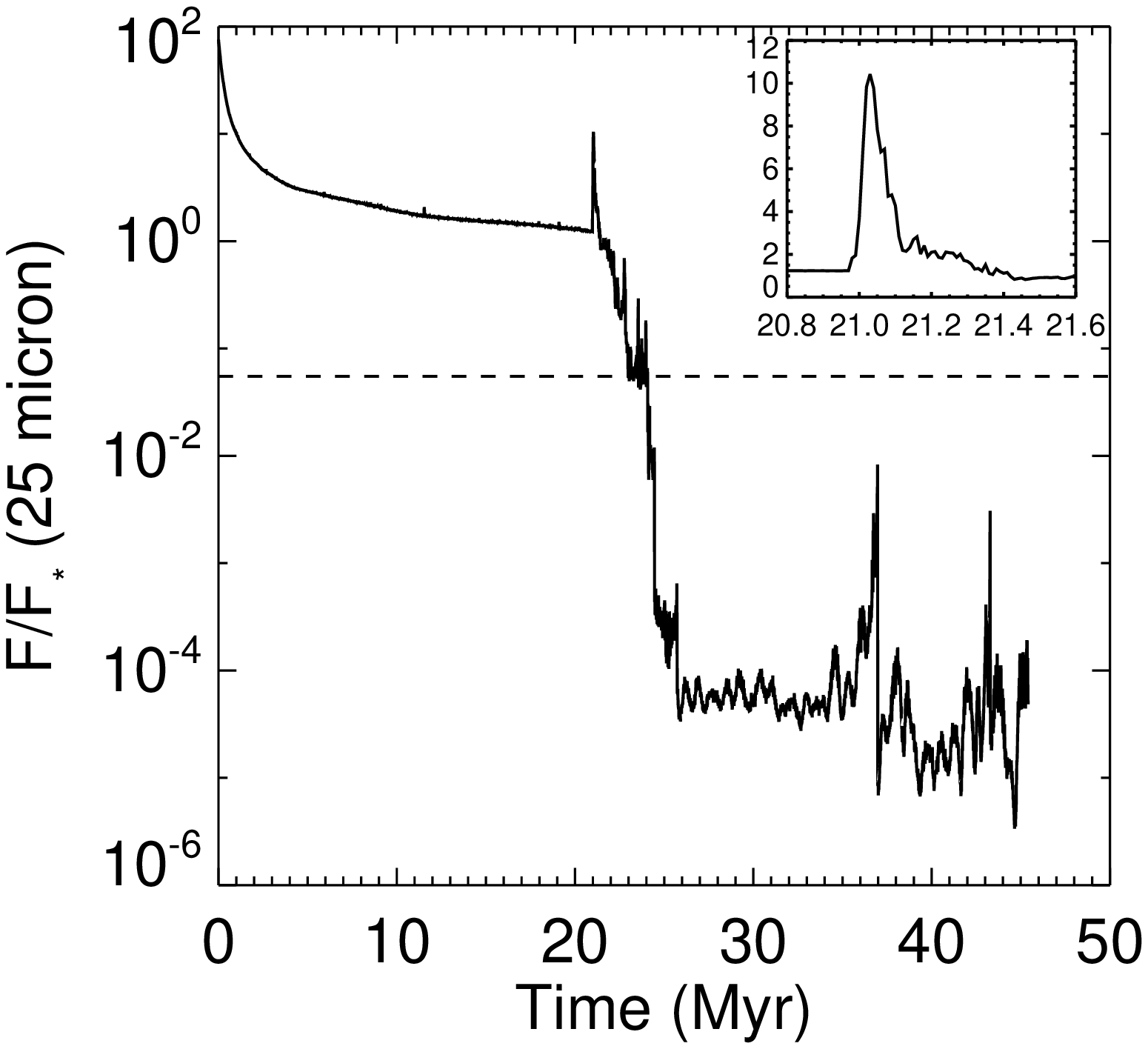}
\vskip .2in
\caption{Evolution of a system with a relatively late (21~Myr) instability among the giant planets. {\bf Left:} Snapshots in time of orbital eccentricity vs. semi-major axis for all particles; vertical bars denote $\sin(i)$ for terrestrial bodies with $M_p > 0.2 \ M_\oplus$ and $i > 10^\circ$. The particle size is proportional to the mass$^{1/3}$, but giant planets (large black circles) are not on this scale.  Colors denote water content, assuming a Solar System-like initial distribution (Raymond et al 2004). The surviving terrestrial planet has a mass of 0.72~$M_\oplus$, a stable orbit within the habitable zone, and a high eccentricity and inclination (and large oscillations in these quantities).   {\bf Top right:} The spectral energy distribution of the dust during five simulation snapshots, showing dramatic evolution during and immediately after the instability.  The dashed line represents the stellar photosphere.  {\bf Bottom right:} The ratio of the dust-to-stellar flux at 25 microns as a function of time, including a zoom during the instability.  The rough observational limit of the {\it MIPS} instrument on NASA's {\it Spitzer Space Telescope} is shown with the dashed line (Trilling et al 2008).  All planetesimal particles were destroyed as of 45 Myr via either collision or ejection. A movie of this simulation is available at http://www.obs.u-bordeaux1.fr/e3arths/raymond/scatterSED.mpg.  }
\label{fig:formation}
\end{figure}

The surviving giant planets in our fiducial simulations match the observed eccentricity distribution and we infer the properties of presumed terrestrial exoplanets (Figure~2).  In 40-70\% of {\em unstable} systems -- depending on the initial giant planet masses -- all terrestrial material is destroyed (Veras \& Armitage 2006).  However, matching the exoplanet orbital distribution requires a contribution of up to $\sim$30-40\% of stable systems(Juric \& Tremaine 2008; Zakamska et al 2011), which invariably form systems with two or more terrestrial planets.  The orbits of surviving giant planets act as a measure of the strength of the instability, and as expected (Levison \& Agnor 2003; Raymond et al 2009) terrestrial planet formation is far less efficient for eccentric giant planets.  

\begin{figure}
\includegraphics[width=0.47\textwidth]{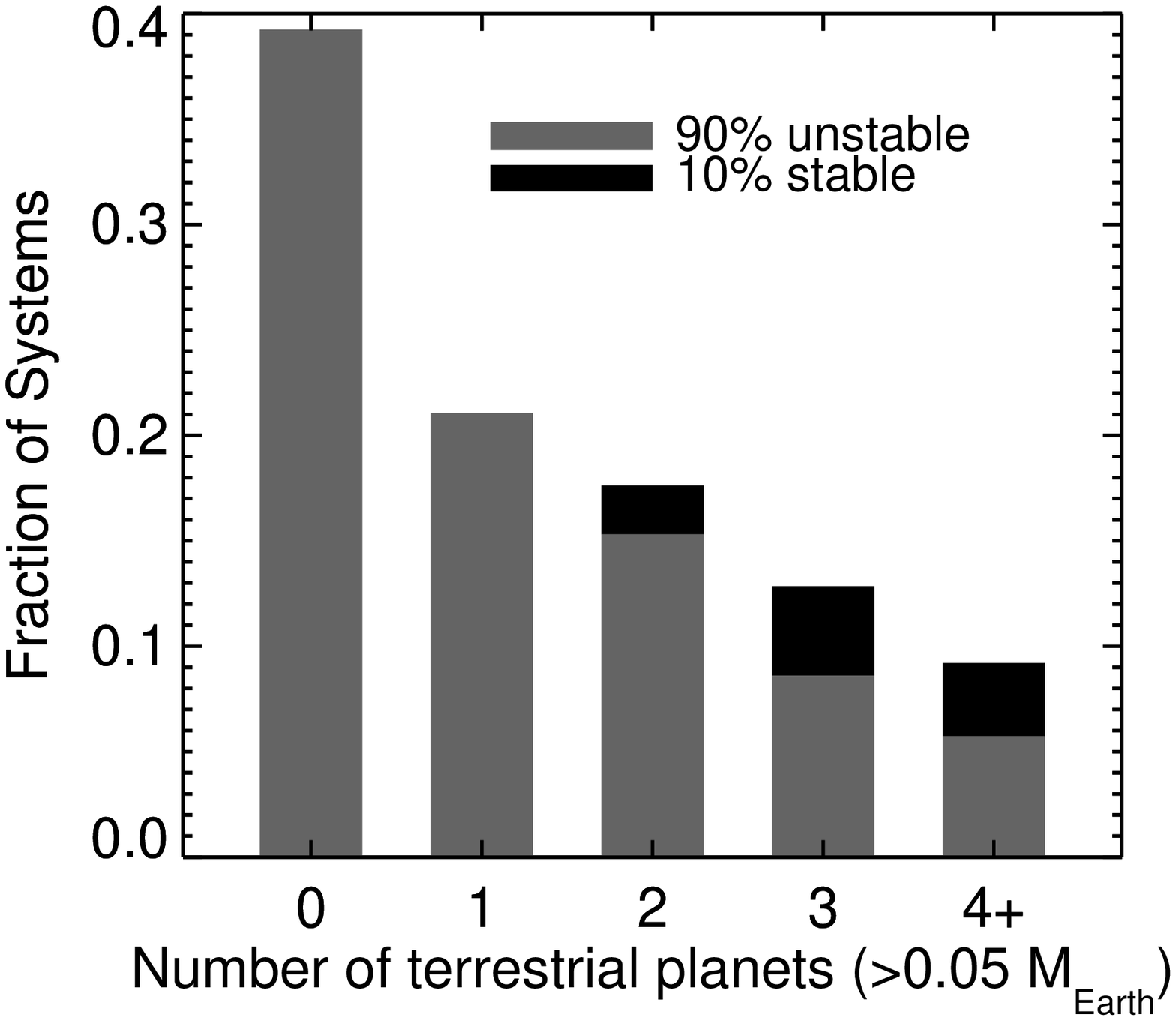}
\includegraphics[width=0.47\textwidth]{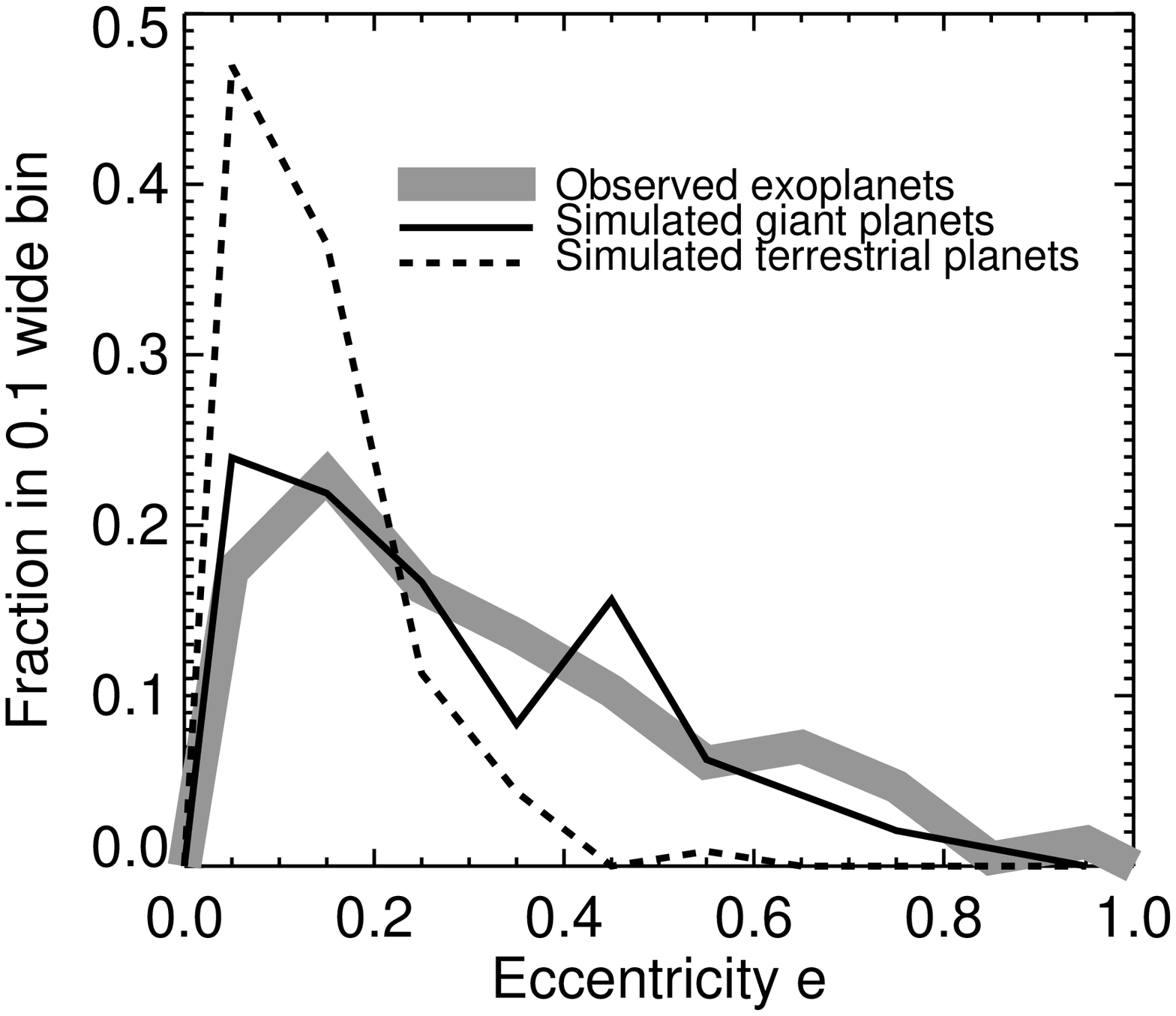}
\centerline{\includegraphics[width=0.47\textwidth]{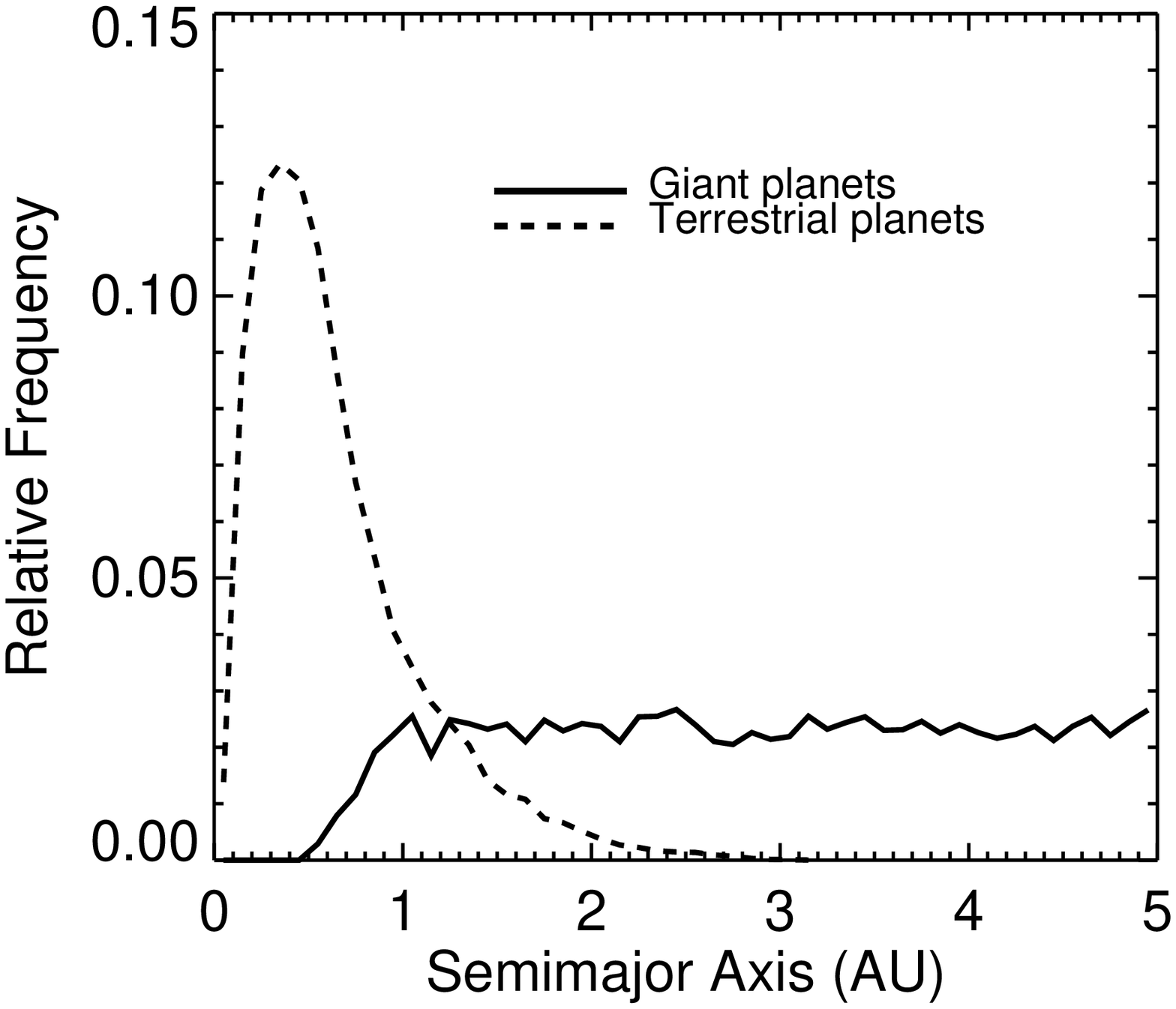}}
\caption{Properties of surviving terrestrial planets.  {\bf Top Left:} Distribution of the number of terrestrial planets ($M_p > 0.05 \ M_\oplus$) in a sample weighted as 90\%  from unstable systems and 10\% from stable systems.  Note that the initial conditions for which the surviving giant planets match the observed exoplanet eccentricity distribution is degenerate between the giant planet masses in a given system and the minority contribution of systems with stable or low-mass giant planets (Raymond et al 2010). 
{\bf Top Right:} Eccentricity distribution of surviving giant and terrestrial planets, compared with the observed giant exoplanets beyond 0.2 AU.  
{\bf Bottom:} Semimajor axis distribution of simulated terrestrial planets (dashed line), derived by scaling the innermost surviving giant planet in each simulation to match an assumed underlying distribution for relevant exoplanets that increases linearly from zero at 0.5 AU and is constant from 1-5 AU.
}
\label{fig:terrestrials}
\end{figure}

As a population, the surviving terrestrial planets have smaller eccentricities than the giant planets (Fig~2; median $e \approx 0.1$ for terrestrials, $0.25$ for giants).  Single terrestrial planets -- formed in 10-20\% of systems -- have somewhat larger eccentricities than in multiple planet systems and undergo much larger oscillations in eccentricity and inclination due to secular forcing from the giant planets.\footnote{These oscillations certainly have important consequences for the planetary climate (Spiegel et al 2010).} If we scale our simulations to match the giant exoplanet semimajor axis-eccentricity distribution beyond 0.5-1 AU (appropriate given the dynamical regime), we can infer the radial distribution of terrestrial exoplanets in the known systems (Fig.~2).  We predict a factor of a few higher abundance of terrestrial planets at a few tenths of an AU than at 1 AU because, given the typical giant-terrestrial planet spacing, planets at 1 AU require distant giant planets that are hard to detect by current methods.  Planets within $\sim 0.1$ AU are sparsely populated because of the assumed inner edge of the embryo disk at 0.5 AU.  

Icy planetesimals -- whose collisional erosion creates debris disks -- survive in dynamically calm environments where the giant planets were either stable, low-mass, or underwent a relatively weak instability.  Indeed, there is a strong anti-correlation between the $70 \ \mu {\rm m}$ dust flux and giant planet eccentricity (Figure~3).  Almost all lower-eccentricity ($e < 0.1-0.2$) giant planets are in systems with debris disks, but at higher eccentricities the fraction of dusty systems decreases as does the dust brightness itself.  There exist some systems with high-eccentricity giant planets and bright dust emission, in agreement with the detected debris disks in known exoplanet systems (Moro-Martin et al 2007).  In these cases the dynamical instability tends to be asymmetric and confined to the inner planetary system and these are therefore not generally good candidates for terrestrial planets, which also agrees with the observed systems (Moro-Martin et al 2010).  The outcome of a given system depends critically on the details of the instability, which is determined by the giant planet masses (Raymond et al 2010).

The total terrestrial planet mass also correlates with debris disk brightness (Fig.~3; the correlation holds for a range of stellar ages and wavelengths).  This correlation exists because the inner and outer planetary system are both subject to the same dynamical environment: the violent instabilities that abort terrestrial planet formation also tend to remove or erode their outer planetesimal disks.  The correlation is not perfect as there exist ''false positives'' with bright dust emission and no terrestrial planets, corresponding to systems with asymmetric, inward-directed giant planet instabilities. Conversely, ''false negatives'' with terrestrial planets but little to no dust are systems that underwent asymmetric but outward-directed instabilities.  

\begin{figure}[t]
\includegraphics[width=0.48\textwidth]{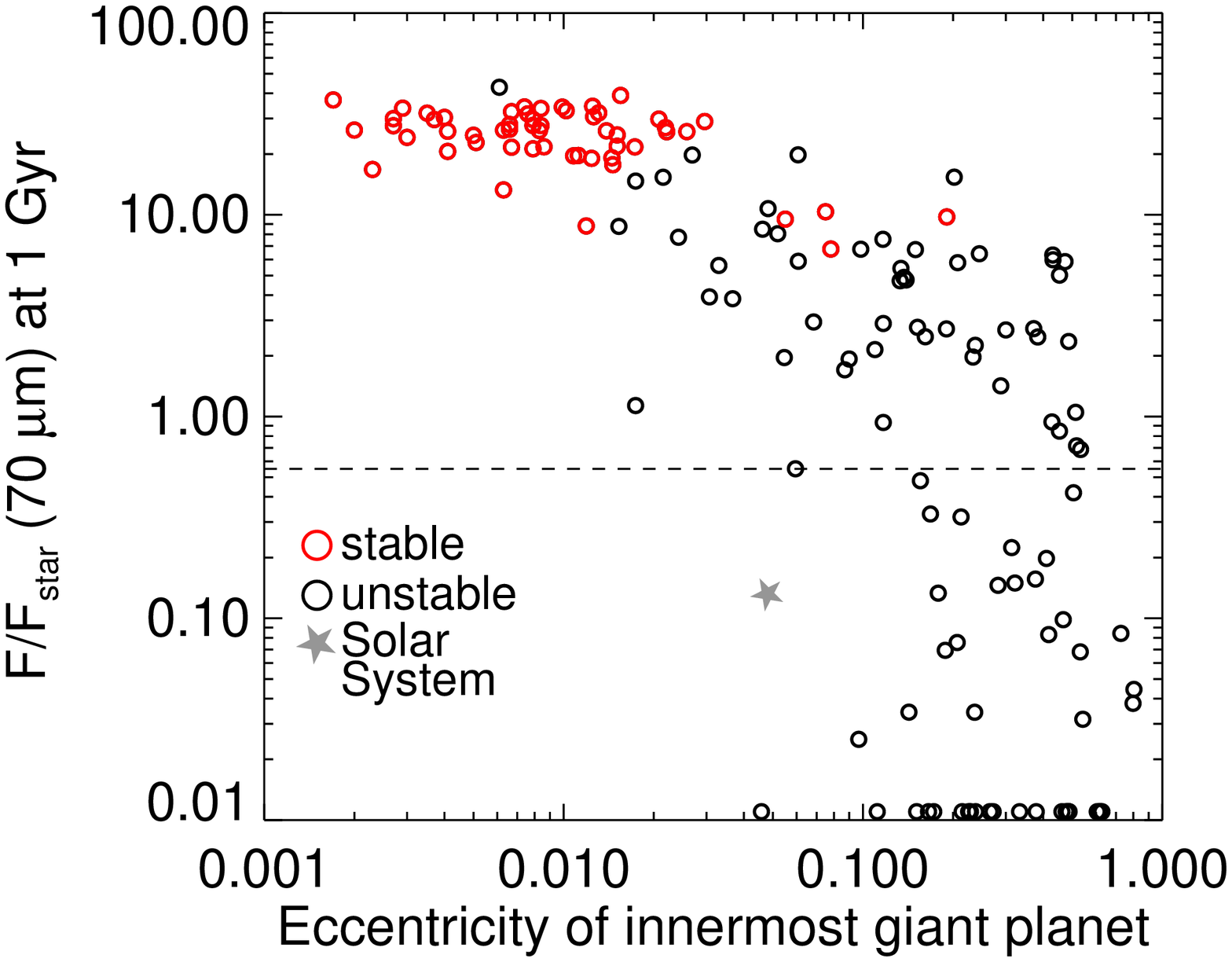}
\hfill
\includegraphics[width=0.48\textwidth]{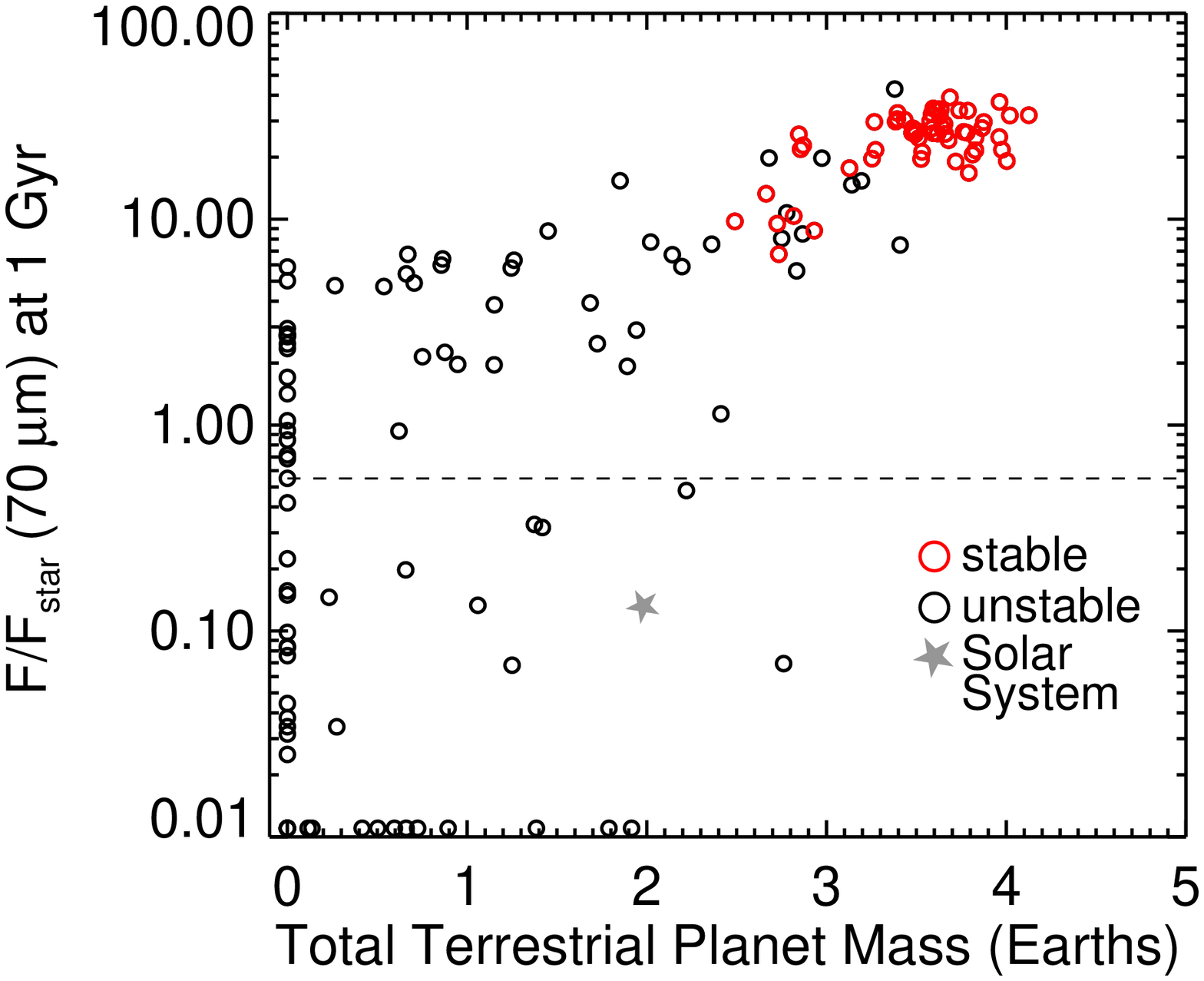}
\caption{Correlations between debris disks and planets. {\bf Left:} The dust-to-stellar flux ratio $F/F_\star$ at 70~$\mu$m after 1~Gyr of dynamical and collisional evolution as a function of the eccentricity of the innermost surviving giant planet.   Unstable systems are in black, stable systems in red, and the dashed line represents an approximate threshold above which excess emission was detectable using {\em Spitzer} data (Trilling et al 2008).  The grey star is an estimate of the Solar System's debris disk flux roughly 900 Myr after the LHB (Booth et al 2009).
{\bf Right:} $F/F_\star$ (70~$\mu$m) vs. the total mass in terrestrial planets.  Debris disks, especially the brightest ones, trace systems that efficiently form terrestrial planets. }
\label{fig:debris}
\end{figure}

\section{Discussion}
The Solar System lies at the very edge of the debris disk correlations in Fig.~3 because of its combination of a rich terrestrial planet system, a low-eccentricity innermost giant planet, and a low dust flux.  To a distant observer, the Solar System's faint debris disk would suggest a dust-clearing instability in the system's past.  However, Jupiter's low-eccentricity orbit would imply that the instability was weak and that the system may in fact be suitable for terrestrial planets.  This naive argument is remarkably consistent with our current picture of the LHB instability as an asymmetric, outward-directed instability that included a scattering event between Jupiter and an ice giant but not between Jupiter and Saturn (Morbidelli et al 2010).  

The observed statistics of debris disks show that 15-20\% of solar-type stars have bright dust emission at $70 \ \mu {\rm m}$ (Trilling et al 2008; Carpenter et al 2009).  Our simulations show that debris disk systems generally represent dynamically calm environments that should have been conducive to efficient terrestrial accretion, and are therefore likely to contain systems of terrestrial planets.  The candidate selection for terrestrial planet systems can be further improved by choosing systems with low-eccentricity giant planets and/or very bright dust emission (e.g., $F/F_{star} \geq 10$ at $70 \ \mu {\rm m}$).  Given the existence of false positives and negatives, the significant uncertainties in our initial conditions, and the existence of other potentially important processes, a detailed statistical analysis is needed to use our simulations to estimate the frequency of debris disks into a robust estimate of the fraction of stars that will be found to harbor terrestrial planets ($\eta_{Earth}$ from the famous Drake equation; see Raymond et al 2011a,b).  Nonetheless, we note that the frequency of $70 \ \mu {\rm m}$ debris disks (15-20\%) is very close to estimates of the frequency of close-in, few Earth-mass planets (10-30\% -- Howard et al 2010; Mayor et al 2009) and our simulations suggest that this is not a coincidence but a natural outcome of planet formation.

\vskip .2in
This proceeding is based on Raymond et al (2011a,b).  The astute reader will notice that only half of S.N.R.'s talk from the IAU is covered here; the other half of the talk, which dealt with a new model for Solar System formation, is under embargo (Walsh et al 2011).  Simulations were run at Weber State University and at Purdue University (supported in part by the NSF through TeraGrid resources).  S.N.R. thanks CNRS's PNP and EPOV programs and NASA Astrobiology Institute's Virtual Planetary Laboratory lead team for funding. P.J.A. acknowledges funding from NASA and the NSF.

\end{document}